\documentstyle [preprint,aps]{revtex}
\begin{document}
\preprint{MSUCL-924}
\title{On the mean free paths of pions and kaons in hot hadronic matter}
\author{Kevin Haglin and Scott Pratt}
\address{National Superconducting Cyclotron Laboratory and
Department of Physics and Astronomy\\
Michigan State University, East Lansing MI  48824}
\maketitle
\begin{abstract}
\baselineskip=14pt

Pion and kaon mean free paths within a thermal hadronic background are
calculated using relativistic kinetic theory.  Free cross sections are used
which include contributions from $\rho$, $K^{*}$, $\Delta$ and heavier
resonances.  Given pion to baryon ratios appropriate for the breakup stage of a
$200\cdot A$GeV relativistic-heavy-ion collision, we find for temperatures less
than 100 MeV, kaons have a shorter mean free path than pions, while for higher
breakup temperatures the reverse is true.  Since breakup temperatures should be
in the neighborhood of 100 MeV, this suggests that kaon interferometry samples
the same emission distribution as pion interferometry.

\end{abstract}
\baselineskip=14pt

\vskip 2 \baselineskip

By measuring outgoing hadrons in a heavy-ion collision one can reconstruct a
reaction's breakup stage.  The momenta of outgoing particles can be
measured directly and a space-time picture of the last collisions may be
constructed using the techniques of two-particle interferometry\cite{PrattQM}.
Space-time information is especially useful.  For instance, if the time a
reaction takes to proceed is much longer than 10 fm/c, it would signal a
reduction in pressure, inferring a first-order phase transition\cite{Pratt86}.

Unfortunately, such information regarding the reaction's lifetime can be
masked by the presence of long-lived resonances\cite{Padula}, particularly the
$\omega$ which has a lifetime of 20 fm/c.  For this reason Padula and Gyulassy
campaigned for two-kaon interferometry.  The only long-lived resonance
responsible for a significant portion of kaons is the $K^*$(892) which has a
lifetime of 8 fm/c, smaller than characteristic times for crossing the
reaction zone.

In this letter we study mean free paths of pions and kaons at temperatures
and densities characteristic of hadronic matter from the breakup stage of a
relativistic heavy-ion collision at CERN.  We find that in this environment
mean free paths are remarkably similar and conclude that information
from kaon interferometry describes the dissolution of the entire system, as
kaons should escape at the same time as pions which comprise the bulk of
the matter.

In baryon-rich matter, such as is characteristic of heavy-ion collisions at the
AGS, most charged kaons are positive since $s$ quarks are absorbed into
baryons, leaving ${\bar s}$ quarks to form hadrons.  Positive kaons interact
little with baryons as opposed to pions which interact vigorously due to
the $\Delta$ resonance.  Thus positive kaons escape baryonic matter much
more easily than pions and the results of kaon interferometry indeed
reflect this fact\cite{Vossnack}.  For sulfur projectiles at 200 A$\cdot$
GeV incident on heavy targets, resulting pion multiplicities are ten
times as large as baryon multiplicities at mid-rapidity.  Thus a meson's
escape probability depends principally on its cross section with pions.

In the context of a thermodynamic model we estimate mean free paths of a given
meson of type $a$ given its momentum.  Assuming the meson interacts with an
assortment of hadrons of type $b$ whose density is given according to a
relativistic Boltzmann distribution and assuming we know the cross sections
$\sigma_{ab}$, the mean free path is:

\begin{equation}
\lambda_a({\bbox{p}}_a) = \frac{{\bbox{p}}_a}{E_a}
\frac{1}{R_a^{net}({\bbox{p}}_a)}.
\end{equation}

Here ${\bbox{p}}_a$ and $E_a$ represent the meson's momentum and
energy while $R_a$ is
the collision rate.  The net collision rate includes contributions from all
different species.

\begin{eqnarray}
R_a^{net}({\bbox{p}}_a) &=& \sum_b R_{ab}({\bbox{p}}_a)\\
\nonumber R_{ab}({\bbox{p}}_a) &=& \int ds \frac{d^3p_b}{(2\pi )^3}
f({\bbox{p}}_b)
\sigma_{ab}(s) v_{rel} \delta \left(s-(p_a +p_b)^2\right)
\label{eq:rate}
\end{eqnarray}
where
$$
v_{rel} = \frac{\sqrt{(p_a\cdot p_b)^2 - m_a^2m_b^2}}{E_aE_b},\hspace*{20pt}
f({\bbox{p}}_b) = (2s_b +1) e^{-(E_b-\mu )/T}.
$$

Only baryons are given a chemical potential which is chosen to result in a
free $\pi^{+}$ to proton ratio of 10.  Relative densities of
hadron species are dependent on the temperature.  They are illustrated
in Fig. 1.  Species included in the calculation are $\pi$, $K$, $\rho$,
$\eta$, $\omega$ and $K^*$.

Cross-sections are dominated by contributions of resonances.  For pions the
largest contributor to the collision rate is the resonant reaction through the
$\rho$ while for kaons the most common collision is with pions through the
$K^*$ resonance.  Resonant cross-sections are assumed to have the form:

\begin{equation}
\sigma_{ab}(s) \propto \frac{s-(m_a - m_b)^2}{(s-m_R^2)^2 +
m_R^2\Gamma_R^2}
\end{equation}

The mass and width of the resonance are $m_R$ and $\Gamma_R$ respectively.
Normalizations to this form are chosen to yield maximum cross sections
consistent with unitarity limits.  The unitarity limits for cross sections
through a resonance are proportional to $1/k^2$ where $k$ is the relative
momentum of the scattering.

For a narrow width $\Gamma_{R}$ the contribution to the rate in
Eq.~(\ref{eq:rate}) is proportional to $\bar{f}(k_{R})\Gamma_{R}$,
where $k_{R}$ is the magnitude of the reduced relative momentum at the
resonance and $\bar{f}$ is the average Boltzmann factor for a particle
$b$ having a momentum corresponding to a resonance with $a$.
One expects $\pi\pi$ scattering to be weaker than $\pi K$ because
the $K^*$ has less decay energy than a $\rho$, hence the
Boltzmann factor is larger for the $K^*$ case.  However,
the width of the $K^*$, 50 MeV, is one third the width of the $\rho$.  At
sufficiently high temperatures pions scatter more readily when travelling
through a pion gas than do kaons.

Fig. 2 shows the mean free paths of particles travelling through a hadron gas
at a temperature of 120 MeV as a function of their momentum. The dotted lines
show the mean free paths of a pion (Fig. 2a) and a kaon (Fig. 2b) in a pion gas
where only interactions through the $\rho$ and $K^*$ resonances are
considered.  These results are consistent with previous calculations
which used similar assumptions\cite{jlgoity89,prakash93}.
The mean free paths go to zero with zero momentum due to the velocity term in
Eq. (1). The rise of the mean free path for pions at low momentum demonstrates
how pions without much energy have difficulty colliding through the $\rho$
resonance since it is difficult to find a second pion with sufficient
energy to produce a $\rho$.  For kaons this is not so difficult, since
$K^*$\,s do not require so much center-of-mass energy; and since the
kaon is heavier, less of a colliding particle's energy is lost to
center-of-mass motion.

The long-dashed curves result from adding to the
resonant $\rho$ and $K^{*}$ an s-wave contribution.  Boltzmann weighting
favors near-threshold interactions and therefore enhances the effect
of the relatively small s-wave for which a constant cross section of
8.0 mb is taken.  Next we add the heavier resonances.
For the pion results (Fig. 2a) we add $K^*$, $a_{1}(1260)$ and
$\Delta$ for pion-nucleon interactions.  They are presented as
short-dashed, dot-dashed and solid lines, respectively.  We also studied
the effect of even heavier resonances but conclude they are of no importance
for these temperatures.  The effect of the resonances beyond $\rho$ is
to reduce the mean free path
of order 30\% at low pion momentum and 15\% at high
momentum.

To the kaon interactions we add $\phi$ and $K_{1}(1270)$
for interactions with other kaons and with $\rho$\,s.  We also add
$\Lambda(1520)$ and $\Lambda(1800)$ for $K^{-, \bar{0}}$\,s interacting
with nucleons.
The results are again shown as long-dashed, short-dashed, dot-dashed and
solid lines, respectively.  The effect of the $\phi$ is quite small
as expected.  The $K_{1}(1270)$ is near enough to threshold that the
unitarity limit is relatively large resulting in a rather strong
interaction and a noticeable change in the kaon mean free path.
Finally, the presence of nucleons further reduces the kaon mean
free path but only slightly.  Overall these heavier resonances
play a modest role.  On the other hand, quantitative
comparisons between experimental data and model calculations
that neglect these heavier resonances such as refs.~\cite{sorge,pang,csernai}
are only reliable, as we have shown, up to tens of percents.
Many authors have studied mean collision times for
pions\cite{sg85,es88,pl91,mp93},
but we choose to compute average mean free paths of pions and
compare to kaons.  The result is shown in Fig. 3
as a function of the temperature.  The average mean free path is

\begin{equation}
{\bar \lambda} = \frac{\int d^3p f({\bbox{p}}) \lambda ({\bbox{p}})}{\int d^3p
f({\bbox{p}})}.
\end{equation}

All resonances and s-wave contributions are
included in the calculation.  The reaction should end when the mean free path
is near the size of the system.  From interferometry the size of the
dissolving system appears similar to the size of a lead nucleus which has
a radius of seven fm.  A mean free path of seven fermi corresponds to a
temperature of approximately 110 MeV.  One must be cautious of such
conclusions both because of the lack of geometric detail involved in the
inference and the questionable assumption of zero chemical potential.  Rapid
expansion can outrun a system's ability to stay in chemical equilibrium,
resulting in large chemical potentials\cite{Gavin,Welke,Gerber},
higher densities and therefore shorter mean free paths.  This would
allow the system to stay together longer, resulting in lower breakup
temperatures.

The most remarkable aspect of Fig. 3 is that the mean free paths of kaons and
pions are so similar.  This means that kaons can be used to view the final
stage of the collision without the qualification that they have escaped
prematurely.  This does not mean that correlation functions from kaons and
pions should have the same apparent source sizes, and indeed at CERN
preliminary measurements point to smaller sizes for kaons than
pions\cite{Humanic}.  Even if one can account
for pions from long-lived resonances, kaon sources can appear smaller due to
collective expansion.  A heavier particle with a given velocity is more
confined to the region with the same collective
velocity\cite{Pratt84,Pratt86}.  Given that pions and kaons have such
similar escape probabilities, one can then compare and interpret
correlation results from kaons and pions.  This clarifies the meaning of
both measurements.  In addition the effect of heavier resonances and
s-wave contributions has been cataloged, showing they matter at the level
of 10--20 percent.  These calculations can provide guidance to
those constructing transport models.  By using the
results of Fig. 2, one can see which channels should be included
and which reactions can safely be neglected.

{\acknowledgements  {This work was supported by W. Bauer's Presidential Faculty
Fellow Award, NSF grant no. PHY-925355, and from NSF grant no. PHY-9017077.}}

\begin{figure}
\caption{Thermal ensemble of hadrons and their number densities
as a function of temperature.}
\label{fig:one}
\end{figure}
\begin{figure}
\caption{Mean free paths of pions (a) and kaons (b) as a function of
their momentum.  Resonant $\rho$ and $\protect K^*$
cross sections are included in the dotted curves.  The the effects of
adding to this the s-wave (long-dashed curves) into the cross section
are shown.  Finally, $K^*$, $a_{1}(1260)$ and $\Delta$ into the pion
results and $\phi$, $K_{1}(1270)$, $\Lambda(1520)$ and $\Lambda(1800)$
are shown as short-dashed, dot-dashed and solid lines, respectively.}
\label{fig:two}
\end{figure}
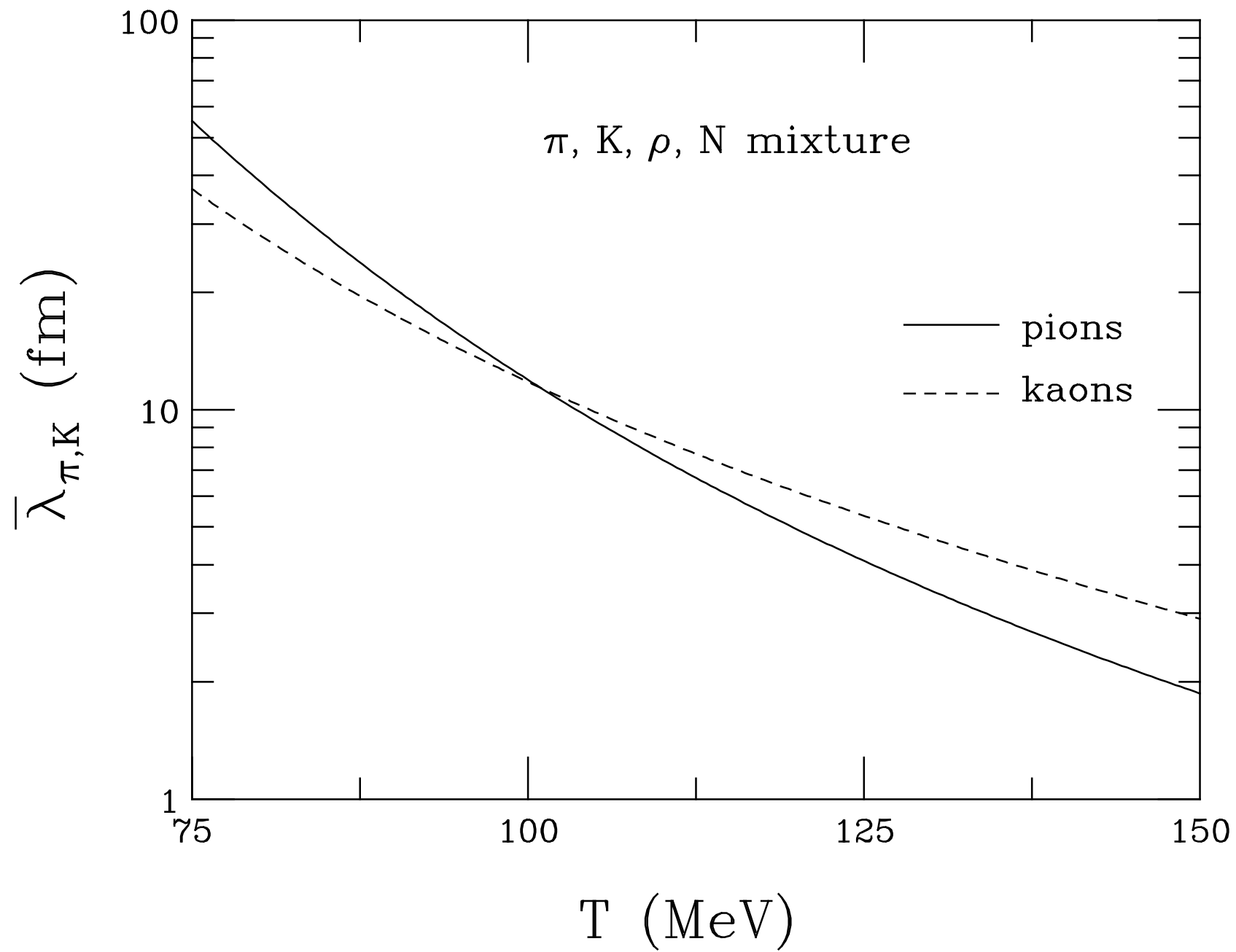
\begin{figure}
\caption{Average mean free paths of pions (solid curve) and kaons (dashed
curve) as they depend on the temperature.}
\label{fig:three}
\end{figure}


\begin{references}

%
\bibitem{PrattQM}S. Pratt {\em et al.}, Nucl. Phys. A{\bf 566} (1994) 103c.
\bibitem{Pratt86}S. Pratt, Phys. Rev. D {\bf 33} (1986) 1314.
\bibitem{Padula}M. Gyulassy and S. S. Padula, Phys. Rev. D {\bf 41} (1990) R21.
\bibitem{Vossnack}O. Vossnack, HIPAGS Proceedings, ed. G. S. F. Stephans,
S. G. Steadman and W. L. Kehue, MITLNS-2158 (1993).
\bibitem{jlgoity89}J. L. Goity and H. Leutwyler, Phys. Lett. {\bf B} 228 (1989)
517.
\bibitem{prakash93}M. Prakash, M. Prakash, R. Venugopalan and
G. Welke, Phys. Rep. {\bf 227} (1993) 321.
\bibitem{sorge}H. Sorge, R. Mattielo, A. Jahns, H. St\"ocher and W.
Greiner, Phys. Lett. {\bf B}271 (1991) 37.
\bibitem{pang}Y. Pang, T. J. Schlagel and S. H. Kahana, Phys. Rev. Lett.
{\bf 68} (1992) 2743.
\bibitem{csernai}L. V. Bravina, N.S. Amelin, L. P. Csernai, P. Levai and
D. Strottman, Nucl. Phys. A{\bf 566} (1994) 461c.
\bibitem{sg85}S. Gavin, Nucl. Phys. A{\bf 435} (1985) 826.
\bibitem{es88}E. V. Shuryak, Phys. Lett. {\bf B}207 (1988) 345.
\bibitem{pl91}P. Levai and B. M\"uller, Phys. Rev. Lett. {\bf 67} (1991)
1519.
\bibitem{mp93}M. Prakash, M. Prakash, R. Venugopalan and
G. Welke, Phys. Rev. Lett. {\bf 70} (1993) 1228.
\bibitem{Gavin}S. Gavin and P. V. Ruuskanen, Phys. Lett. {\bf B}262
(1991) 326.
\bibitem{Welke}G. M. Welke and G. F. Bertsch, Phys. Rev. C {\bf 45} (1992)
1403.
\bibitem{Gerber}P. Gerber, H. Leutwyler and J. L. Goity, Phys. Lett.
{\bf B}243 (1990) 18.
\bibitem{Humanic}T. J. Humanic (NA44 Collaboration), Nucl. Phys.
A{\bf 566} (1994) 115c.
\bibitem{Pratt84}S. Pratt, Phys. Rev. Lett. {\bf 53} (1984) 1219.

\end{references}
\end{document}